# Magnetic order in Tb and Dy substituted $La_3Ni_2O_7$ and effects of high-pressure high-temperature $O_2$ annealing

Yuhang Deng[1], Farhad Taraporevala[1], Jui-Chia Lee[1], Zhekai Yang[1, †], Gabriel Mee[1], Swadhin Rout[1], Richard Erber[1], Fábio Abud[2], Renato F. Jardim[3], Maxim Avdeev[4, 5], Jeffrey W. Lynn[6, 7], M. Brian Maple[1, *]

1. Department of Physics, University of California San Diego, La Jolla, California 92903, USA
2. Center for Human and Natural Sciences, Federal University of ABC, Santo André 09210-580, São Paulo, Brazil
3. Instituto de Física, Universidade de São Paulo, São Paulo 05508-090, São Paulo, Brazil
4. Australian Nuclear Science and Technology Organisation, Sydney, New South Wales 2234, Australia
5. School of Chemistry, The University of Sydney, Sydney, New South Wales 2006, Australia
6. NIST Center for Neutron Research, National Institute of Standards and Technology, Gaithersburg, MD 20899, USA.
7. Quantum Materials Center, Department of Physics, University of Maryland, College Park, MD 20742, USA

[†]Present address: Department of Physics, The Hong Kong University of Science and Technology, Hong Kong, China

[*]Corresponding author: M. Brian Maple, mbmaple@ucsd.edu

## Abstract

To study the influence of magnetic lanthanide (*Ln*) substituents on $La_3Ni_2O_7$, polycrystalline samples of $La_{3-x}Ln_xNi_2O_7$ (*Ln* = Pr, Eu, Gd, Tb, Dy, Ho, Er, Tm, and Yb) were synthesized and characterized by electrical transport, magnetization, and x-ray and neutron powder diffraction before and after high-pressure high-temperature oxygen annealing. The annealing process led to the formation of $La_3Ni_2O_7$ polymorphs and more metallic-like temperature dependence of the electrical resistance while changing the density-wave transition temperature only slightly. For unannealed $La_{2.7}Tb_{0.3}Ni_2O_7$ and $La_{2.7}Dy_{0.3}Ni_2O_7$, distinct features in the magnetic susceptibility and additional magnetic peaks in neutron diffraction patterns indicate that both samples develop long-range magnetic order below ~13 and 11 K, respectively. The magnetic signatures were strongly suppressed following oxygen annealing, suggesting a possible association with the accompanying structural modifications. These behaviors distinguish Tb and Dy from Pr, Nd, and Sm substituents which do not introduce magnetic ordering in the system, establishing $La_{3-x}Tb_xNi_2O_7$ and $La_{3-x}Dy_xNi_2O_7$ as promising platforms for exploring the interplay between magnetic order and pressure-induced superconductivity in $La_3Ni_2O_7$.

Bilayer nickelate, magnetic ordering, density wave transition, metallization, oxygen annealing

## 1. Introduction

High temperature superconductivity with critical temperatures ($T_c$) in the liquid nitrogen temperature range has been discovered in several material families, including cuprates such as $YBa_2Cu_3O_{7-\delta}$ (YBCO) [1, 2], hydrogen-rich sulfide [3] and metal hydrides [4, 5] under megabar pressures. Nickelates with Ruddlesden-Popper (RP) structures [6, 7] which contain $NiO_2$ planes comprised of corner-sharing Ni-O octahedra closely resembling the $CuO_2$ planes in YBCO and other high $T_c$ cuprates, have long been considered as promising candidates for high temperature superconductivity [8]. Following the discovery of reduced infinite-layer superconducting $RE_{1-x}AE_xNiO_2$ (RE = rare earth, AE = alkaline earth) thin films [9, 10, 11], RP nickelate superconductors under high pressure have been reported including bilayer $La_3Ni_2O_7$ [12, 13, 14] and trilayer $La_4Ni_3O_{10}$ [15, 16, 17] beginning in 2023. Among them, $La_3Ni_2O_7$ has attracted particular attention because it exhibits a higher $T_c \approx 80$ K above a moderate pressure of 14 GPa. Through precisely controlling the valence of Ni by the floating-zone method in a high-pressure oxygen atmosphere [12, 13], single crystalline metallic $La_3Ni_2O_7$ can be synthesized, exhibiting zero resistivity [13] and the Meissner effect [18, 19] under hydrostatic pressure. More recently, superconductivity has also been reported in thin films of $La_3Ni_2O_7$ [20] and its lanthanide (*Ln*) substituted cousins (e.g., $La_{3-x}Pr_xNi_2O_7$ [21, 22]) at ambient pressure, although with lower $T_c$ values than those of their bulk counterparts. This behavior has been attributed to the compressive strain imposed by the substrate which may partially mimic the effect of external pressure.

The emergence of the high-pressure superconductivity in $La_3Ni_2O_7$ and its *Ln* substituted derivatives is commonly associated with a structural transition from the orthorhombic *Amam* to either the orthorhombic *Fmmm* or the tetragonal *I*4/*mmm* phase [12, 19, 23, 24]. During this transition, the bond angle between Ni-O octahedra from neighboring layers changes from $\theta_{Ni\text{-}O\text{-}Ni} = 168°$ to 180° [12]. However, tetragonal $La_3Ni_2O_7$ with straight interlayer Ni-O-Ni bonds has been synthesized, showing no superconductivity at ambient or high pressure up to 68 GPa, challenging the prevailing view that straightening of the Ni-O octahedra is the key structural ingredient for superconductivity in $La_3Ni_2O_7$ [25]. Instead, the spin density wave (SDW) present in orthorhombic $La_3Ni_2O_7$ but absent in its tetragonal polymorph has been proposed as a prerequisite for the superconductivity [25]. At ~140 K at ambient pressure, a density-wave-like (DW) transition often shows up [13, 14], which occurs at lower temperatures with increasing pressure until it finally vanishes before the appearance of superconductivity. Understanding the nature of this DW transition, e.g., whether it involves an SDW, a charge density wave (CDW), or an intertwined SDW-CDW [26, 27], could provide insight into the superconducting pairing mechanism of $La_3Ni_2O_7$ [28, 29].

The Ni valence, interlayer Ni-O-Ni bond angle, and DW state are key factors governing superconductivity in $La_3Ni_2O_7$. Substituting La with *Ln* ions provides a promising route for tuning these parameters, as the smaller *Ln* ions can introduce chemical pressure and potentially influence the DW state through their localized magnetic moments. To date, Pr [24], Nd [30], and Sm [31] substitutions have been attempted and the resulting $La_{3-x}Ln_xNi_2O_7$ ($x \approx 1$) compounds all exhibit superconductivity under pressure. Surprisingly, these *Ln* substituted nickelates have comparable values of $P_c$ and $T_c$ as those of $La_3Ni_2O_7$, similar $T - P$ phase diagrams [24, 30, 31] and nearly unaltered DW transition temperatures ($T_{DW}$) at ambient pressure [24, 30]. This invariance is also notable because Pr and Sm substitutions reduce $\theta_{Ni\text{-}O\text{-}Ni}$ [24, 31], a structural change generally thought to disfavor superconductivity. This indicates the insignificant effects of localized 4*f* electrons of Pr, Nd, Sm substituents on the electronic and superconducting properties of the host La nickelate. Oxygen annealing provides another route for tuning the electronic state by modifying the occupancy of the inner apical oxygen sites [32] and the Ni valence. It is generally believed that proper high-pressure high-temperature (HPHT) $O_2$ annealing converts insulating O-deficient $La_3Ni_2O_{7-\delta}$ into more metallic $La_3Ni_2O_7$ [32] thus promoting superconductivity. Consistently, a low volume fraction of superconductivity at 80 K at ambient pressure in 100 bar $O_2$-annealed $La_3Ni_2O_7$ was reported [33]. On the other hand, certain HPHT $O_2$ annealing processes can be detrimental to the high-pressure superconductivity by introducing impurity phases [34]. Understanding how *Ln* substitution and oxygen annealing affect the superconducting critical pressure ($P_c$), the normal-state transport properties, and the superconducting phase of $La_3Ni_2O_7$ is therefore of considerable interest.

Introducing magnetic moments into superconductors provides a powerful means of probing superconducting mechanisms [35, 36, 37]. In an isostructural series of metallic *Ln*-based compounds, magnetic ordering temperatures frequently scale with the de Gennes factor dG = $(g_J - 1)^2 J(J + 1)$, where $g_J$ is the Landé g-factor and $J$ is the total angular momentum of the *Ln* ion. The dG factor reaches its peak at Gd and remains substantially large for heavier lanthanides (Tb – Ho). Consequently, magnetic ordering temperatures are generally expected to be higher for Gd-, Tb- and Dy-containing compounds. Likewise, the suppression of superconductivity by exchange scattering from localized magnetic moments often scales with the dG factor, suggesting a stronger influence of the heavier lanthanides on superconductivity. However, for certain *Ln* ions with valence instabilities, namely, Ce, Sm, Eu, Tm, and Yb [36], some striking exceptions to the de Gennes scaling can occur when the hybridization of localized 4*f*- and conduction-electron states is appreciable. This situation can lead to exotic superconducting phenomena such as reentrant superconductivity due to the Kondo effect in superconductors (e.g., $La_{1-x}Ce_xAl_2$, $(La_{1-y}Th_y)_{1-x}Ce_x$ [35]) and magnetic field-induced superconductivity in superconductors (e.g., various Chevrel phase superconductors containing $Eu^{2+}$ substituents due to the exchange field compensation effect [38, 39]). Recently, the Eu-substituted infinite-layer $RE_{1-x}AE_xNiO_2$ compound was found to exhibit field-reentrant superconductivity [40, 41], illustrating the special role of Eu in nickelate systems.

In an effort to explore the effect of *Ln* ions with large magnetic moments on the properties of $La_3Ni_2O_7$ and test theoretical predictions that some $Ln_3Ni_2O_7$ (e.g., $Tb_3Ni_2O_7$ [42]) could exhibit superconductivity at ambient pressure, we synthesized $La_{3-x}Ln_xNi_2O_7$ (*Ln* = Pr, Eu – Yb) polycrystalline samples and studied their electrical transport and magnetic properties before and after treating them in HPHT $O_2$. Previous studies reported low-temperature magnetic anomalies in Tb-substituted $La_3Ni_2O_7$ [43, 44], with different interpretations of their possible origins. In this work, we confirm the presence of magnetic anomalies in $La_{2.7}Tb_{0.3}Ni_2O_7$ and provide neutron powder diffraction evidence for long-range magnetic order (MO), potentially antiferromagnetic (AFM), in both Tb- and Dy-substituted $La_3Ni_2O_7$, with ordering temperatures ($T_N$) of approximately 13 and 11 K, respectively, well below their DW transition temperatures. The two compounds exhibit distinct spin-flop-like transition fields of approximately 1.1 and 0.8 T, respectively. Furthermore, HPHT $O_2$ annealing enhances their electrical conductivity while suppressing the low-temperature magnetic anomalies, possibly associated with structural modifications of the *Amam* phase.

## 2. Experimental methods

*2.1 Sample synthesis*: Polycrystalline $La_2PrNi_2O_7$ and $La_{2.7}Ln_{0.3}Ni_2O_7$ (*Ln* = Eu – Yb) were synthesized by the sol-gel method [24]. The resultant polycrystalline pellets were placed in alumina boats covered with alumina lids and heat treated at ~500 °C under an oxygen pressure of ~70 bar in a homemade furnace for 36 hours, after which the furnace was cooled to room temperature in ~5 hours. The high-pressure high-temperature (HPHT) sample space is provided by a thick-walled (28 mm) INCONEL 610 tube inserted in a tubular Lindberg/Blue M furnace. Details of the setup are described in detail in [45].

*2.2 X-ray diffraction*: Powder X-ray diffraction (PXRD) data were collected using a Bruker D8-Discover diffractometer for $La_2PrNi_2O_7$ before and after annealing and a D8-Advance Eco diffractometer for $La_{2.7}Ln_{0.3}Ni_2O_7$ before and after annealing. Fine powders were obtained by grinding the samples with an agate mortar and pestle. PXRD profiles were obtained in a θ/2θ configuration, using Cu $K_\alpha$ radiation, operating in a 40 kV and 30 mA setting. Typical angular scans ranged between 10 to 100° in 0.02° and 0.05° steps. Rietveld refinements were performed using FullProf suite [46].

*2.3 Electrical resistivity and vibrating sample magnetometry (VSM)*: Electrical resistivity and VSM measurements were made in a Quantum Design DynaCool Physical Property Measurement System (PPMS). The temperature-dependent electrical resistivity $R(T)$ was measured on pelletized samples using a standard four-probe configuration. A low-frequency AC excitation current of ~10 µA was applied to samples mounted on an electrical transport option (ETO) puck with Apiezon® N-grease to ensure good thermal contact between the sample and the puck. VSM measurements including zero-field-cooled (ZFC) and field-cooled (FC) magnetic susceptibility vs temperature, $\chi(T)$, and magnetization vs applied magnetic field, $m(\mu_0 H)$, at selected temperatures were performed using the VSM option.

*2.4 Neutron powder diffraction (NPD)*: Neutron powder diffraction data for $La_{2.7}Tb_{0.3}Ni_2O_7$ and $La_{2.7}Dy_{0.3}Ni_2O_7$, grown in different batches than those for the samples studied in the VSM experiments but with the same magnetic features as revealed by VSM measurements, were collected between 3 and 30 K and at room temperature, and in magnetic fields, using the high-resolution diffractometer Echidna and the high-intensity diffractometer Wombat at the OPAL research reactor, Australian Nuclear Science and Technology Organisation (ANSTO). The diffraction patterns were refined using the Rietveld method implemented in the FullProf software suite. Crystallographic information files (CIFs) were generated from the refined structures, and structural parameters including lattice constants, bond lengths, and bond angles were subsequently calculated using Visualization for Electronic and STructural Analysis (VESTA) [47].

## 3. Results

### *3.1 Structural effects of lanthanide doping and oxygen annealing*

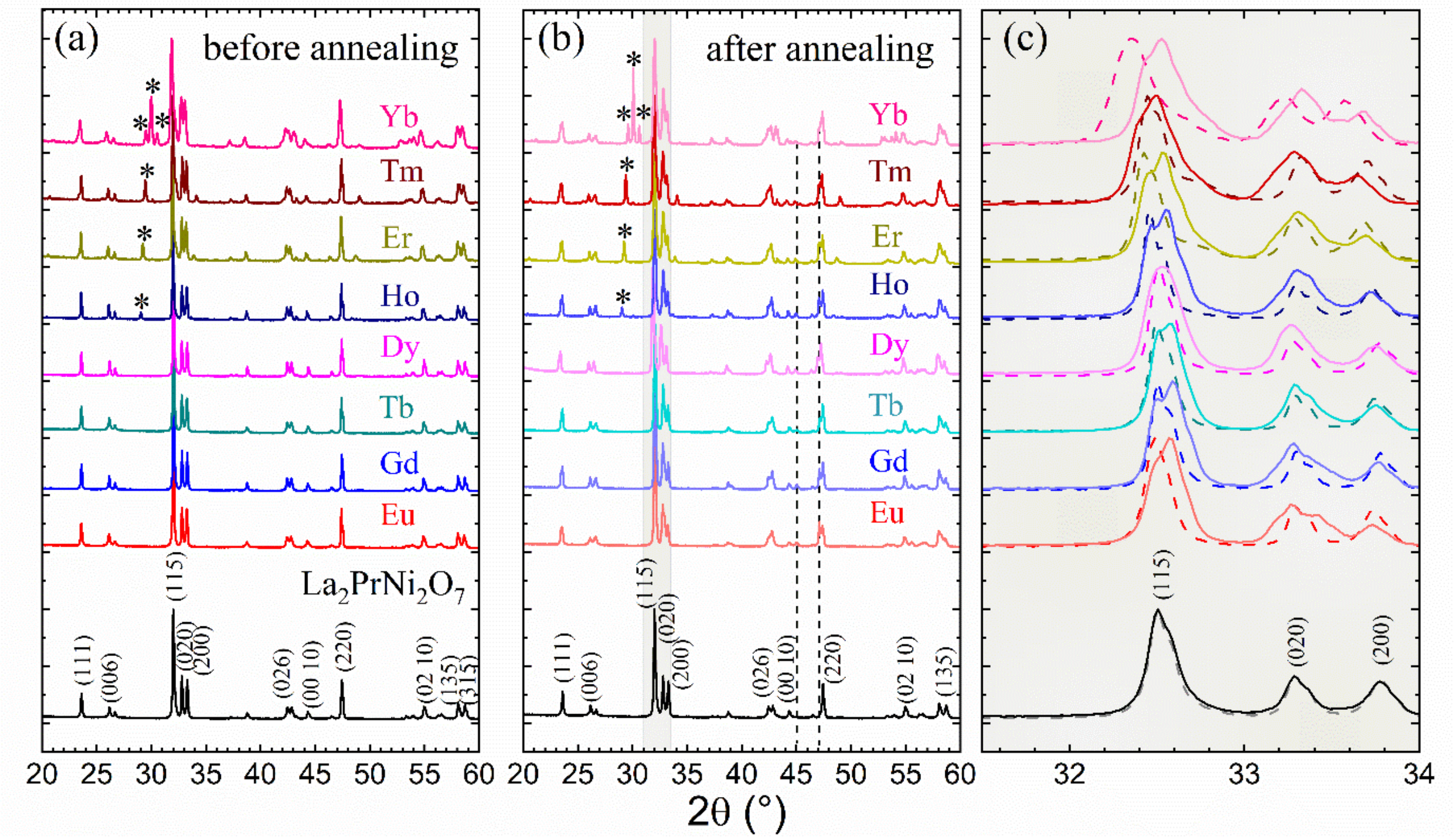

Figure 1. Room temperature x-ray diffraction patterns for $La_2PrNi_2O_7$ and $La_{2.7}Ln_{0.3}Ni_2O_7$ (*Ln* = Eu, Gd, Tb, Dy, Ho, Er, Tm, and Yb) before (a) and after (b) HPHT $O_2$ annealing. The asterisk symbols in (a) and (b) denote the main peaks from impurity phases ($Ho_2O_3$, $Er_2O_3$, $Tm_2O_3$, and $LaYbO_3$). The dashed lines in (b) indicate a new reflection near the (0 0 10) diffraction peak and peak splitting at the (2 2 0) diffraction peak, after annealing. (c) Expanded view of $2\theta \approx 32° - 34°$ region comparing before (dashed) and after (solid) annealing data.

As shown in Fig. 1, before HPHT $O_2$ annealing, Pr, Eu, Gd, Tb, and Dy substituted samples crystallize in the single-phase *Amam* structure [48] with only trace amounts of NiO impurity. Partial substitution of $La^{3+}$ with smaller *Ln* ions (Eu to Dy) progressively contracts the *a* and *c* lattice parameters while leaving the *b* parameter nearly constant, a result of the lanthanide contraction that reduces the unit cell volume by ~ 0.2%. This anisotropic contraction decreases the *a*/*b* ratio, driving the structure toward a more distorted orthorhombic state and away from tetragonal symmetry. In contrast, 10% substitution with heavier lanthanides (Ho, Er, Tm, and Yb) exceeds the solid solubility limit under these synthesis conditions, resulting in multi-phase samples containing secondary oxide phases; the secondary phases, indicated by the asterisks in Fig. 1(a, b), include $Ho_2O_3$ (2.5 wt% or 4.3 mol%), $Er_2O_3$ (7.0 wt% or 11.4 mol%), $Tm_2O_3$ (8.0 wt% or 12.9 mol%), and $LaYbO_3$ (17 wt% or 27 mol%), along with a minor amount of NiO. The progressive emergence of impurities beyond Dy likely reflects a decreasing trend in solid solubility for heavier $Ln^{3+}$ ions due to increasing ionic-size mismatch with $La^{3+}$, in accordance with the Hume-Rothery rules.

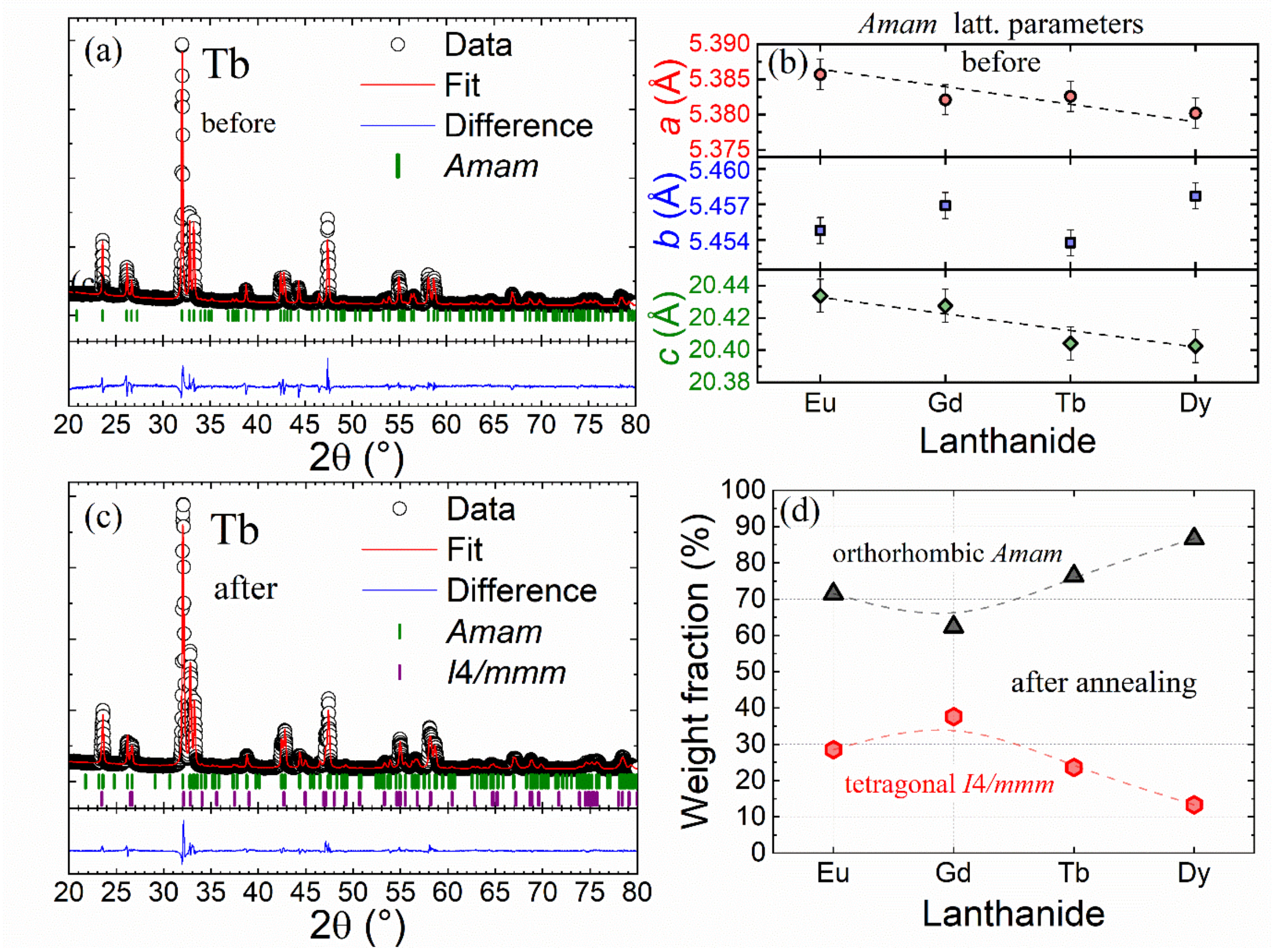

Figure 2. Rietveld refinements of representative $La_{2.7}Tb_{0.3}Ni_2O_7$ data (a) before and (c) after annealing. (b) Evolution of the *Amam* lattice parameters in the Eu – Dy samples prior to annealing. (d) Weight fractions of the *Amam* and *I*4/*mmm* (as a representative secondary structure) phases in the annealed samples.

A primary effect of HPHT annealing is to promote the emergence of a secondary phase in the *Ln*-substituted samples, which, however, does not form in the Pr-substituted specimen under identical treatment. After HPHT annealing, an increase of $Ln_2O_3$ for the Ho, Er, and Tm substituted samples and of $LaYbO_3$ for the Yb substituted sample can be seen. The new reflections in the PXRD patterns (Fig. 1(b, c)) can be indexed to several candidate structural models, including the tetragonal *I*4/*mmm* (2222-type), the orthorhombic *Cmmm* (1313-type), or the higher-symmetry *Fmmm* (2222-type) structures [34, 49]. Among these, the tetragonal structure corresponds to the polymorph reported for unsubstituted $La_3Ni_2O_7$ under HPHT and cooling conditions similar to those used in this work [25, 34, 50, 51]. However, because these candidate models capture the observed diffraction features with similar goodness-of-fit parameters, the secondary phase cannot be unambiguously determined solely from our PXRD data. Nevertheless, using the tetragonal *I*4/*mmm* phase as a representative model (Fig. 2(c)), we demonstrate that the model-dependent secondary phase weight fraction varies across the series, reaching a maximum of nearly 40 wt.% in the Gd-substituted sample (Fig. 2(d)).

Even though NiO, $Ho_2O_3$, $Er_2O_3$, $Tm_2O_3$, and $LaYbO_3$ secondary phases [52] are insulators, they do not appear to have a major effect on the behavior of $R(T)$ for both the unannealed and annealed Ho, Er, Tm, and Yb substituted samples whose $R(T)$ curves have similar shapes as shown in Fig. 7. However, greater caution is required when interpreting the $\chi(T)$ data for these samples, because a significant fraction of *Ln* ions (approximately 23% – 55% and even higher after annealing) resides in impurity phases rather than in the *Amam* $La_3Ni_2O_7$ lattice. Since several of these impurity phases undergo magnetic ordering transitions, their magnetic contributions can complicate the identification and interpretation of the intrinsic magnetic properties of $La_{2.7}Ln_{0.3}Ni_2O_7$.

### *3.2 Magnetic properties of Pr, Eu, Gd, Ho, Er, Tm substituted nickelates*

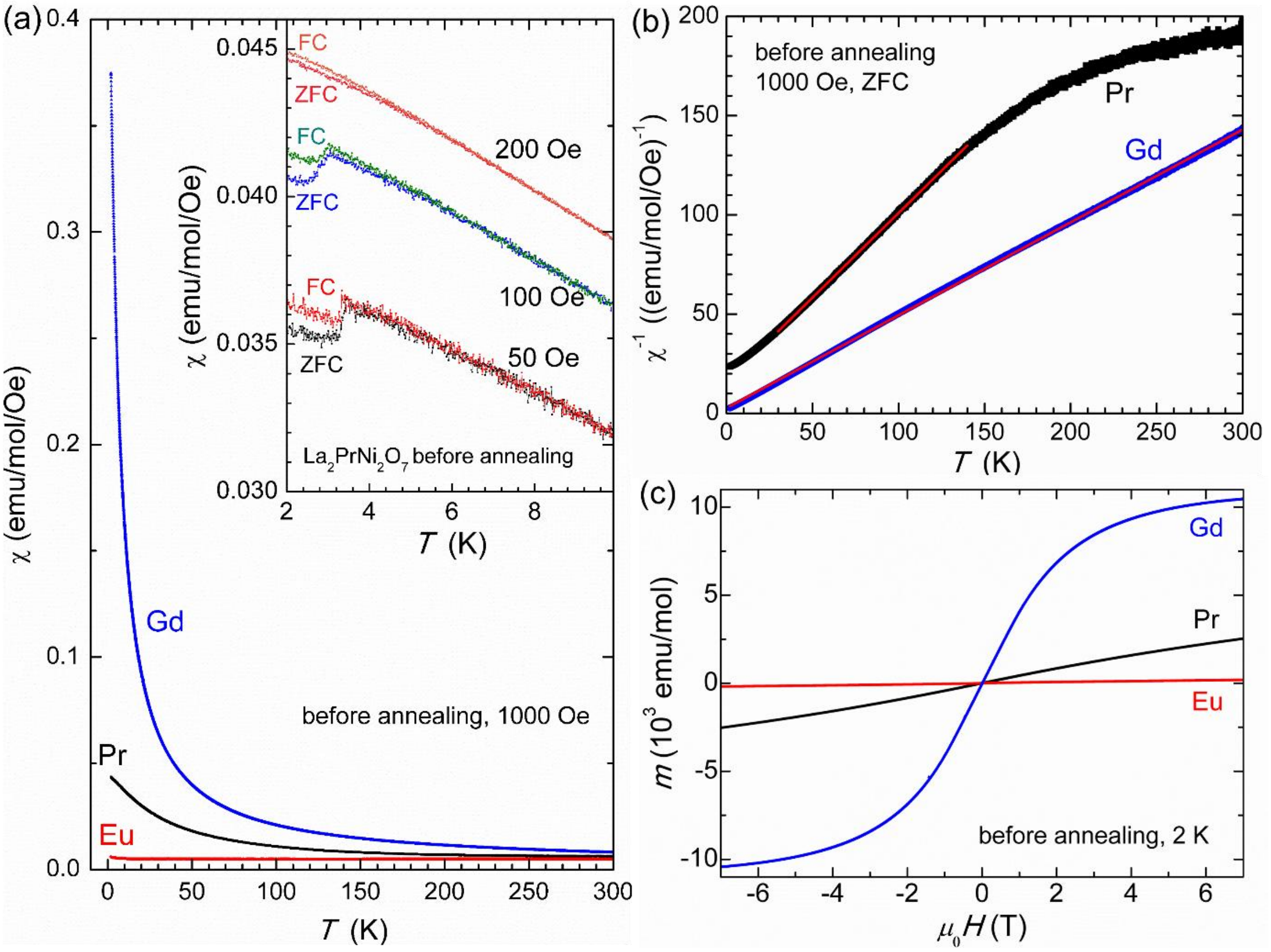

Figure 3. Magnetization of the Pr, Eu, and Gd substituted samples in magnetic fields and at low temperatures before annealing. (a) $\chi$ vs $T$ with an inset showing the anomalous signal of the Pr sample in lower magnetic fields. (b) Background subtracted $\chi^{-1}$ vs $T$ with linear fitting in different $T$ ranges. (c) $m$ vs $\mu_0 H$ at 2 K. Note: 1 emu = $10^{-3}$ A m$^2$, and 1 Oe = $(1000/4\pi)$ A/m.

Magnetization measurements before annealing are summarized in Figs. 3 and 4. For the Pr and Gd substituted samples, $\chi(T)$ after subtraction of a temperature independent background $\chi_0$ = 0.001 emu/mol/Oe [30, 32] coming from $La_3Ni_2O_7$, follow a Curie-Weiss law quite well within certain temperature ranges (particularly for the Gd sample, from 2 K to 300 K), as demonstrated in the linear regions in the $\chi^{-1}(T)$ curves in Fig. 3(b). No low-temperature anomalies indicative of magnetic ordering were observed, except for a small shoulder near 3.2 K in the Er sample, which could be from AFM ordering of an $Er_2O_3$ impurity phase at 3.4 K [53]. The Gd sample exhibits typical paramagnetic behavior in $\chi(T)$ and $m(H)$, suggesting the absence of magnetic order down to 2 K. In contrast, the Eu sample shows a very small and nearly temperature-independent magnetic susceptibility, which indicates Eu ions are in the trivalent $J = 0$ ground state with $\mu_{eff} = 0$. The results for the Ho, Er, and Tm substituted samples (in Fig. S1) and more details about our Curie-Weiss analysis (in Table S1) can be found in the Supporting Information.

A noteworthy feature is observed in χ(*T*) of as-synthesized $La_2PrNi_2O_7$. Although the magnetic response is predominantly paramagnetic, a weak anomaly appears near 3.4 K, evidenced by a bifurcation between the zero-field-cooled (ZFC) and field-cooled (FC) curves and a diamagnetic-like contribution. This feature is readily suppressed by a relatively small applied magnetic field of 200 Oe. While this behavior may be indicative of a superconducting transition with a very small superconducting volume fraction, the present data alone does not provide sufficient evidence to establish superconductivity. Nevertheless, the observation is intriguing considering a previous report of a low superconducting volume fraction with an onset temperature near 80 K at ambient pressure in $La_3Ni_2O_7$ annealed under 100 bar $O_2$ [33].

### *3.3 Magnetic ordering in Tb and Dy substituted nickelates*

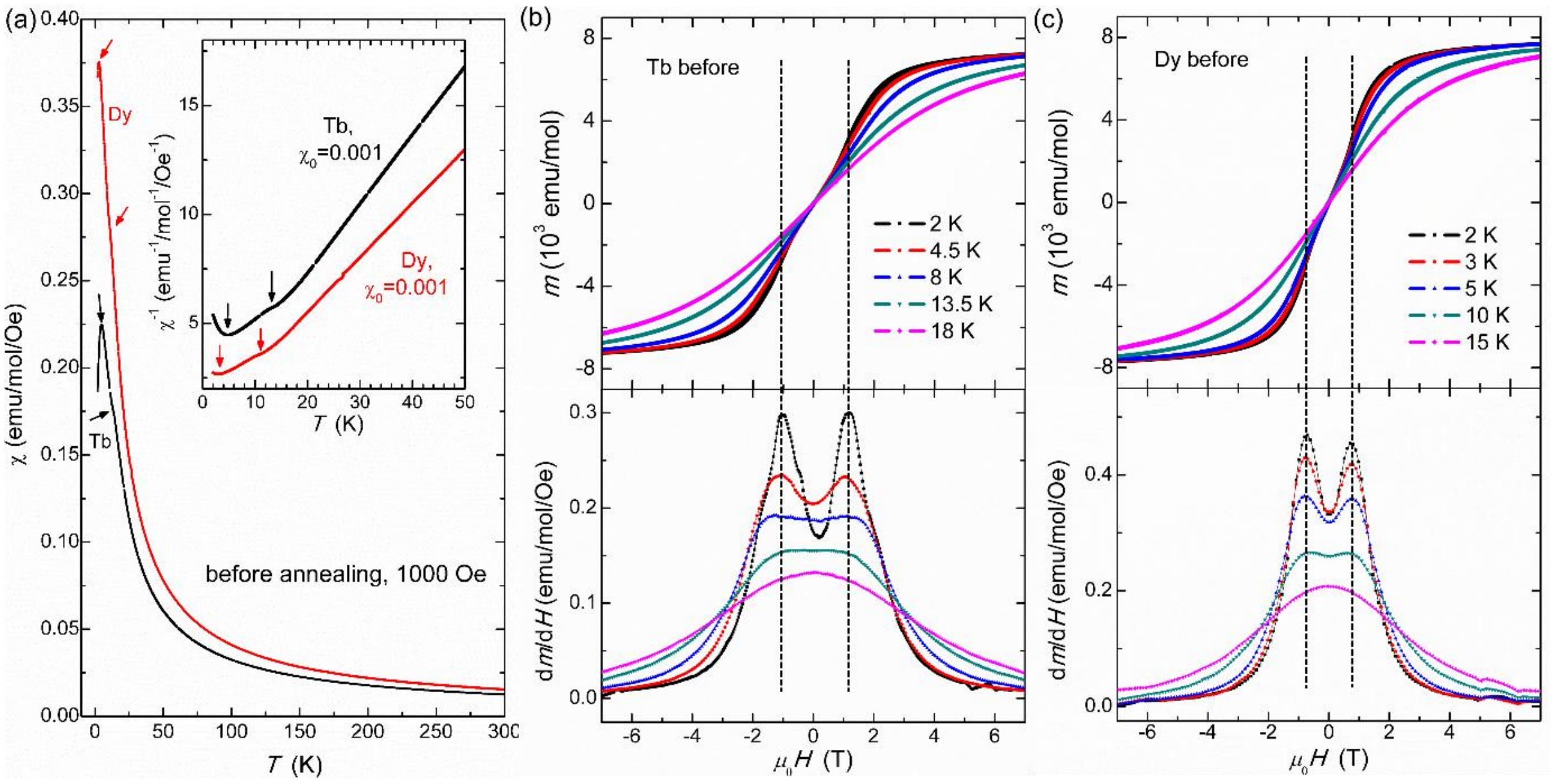

Figure 4. Magnetization of the Tb and Dy substituted samples before annealing in magnetic fields and at low temperatures. (a) χ vs *T* with arrows pointing to peaks or shoulders, with an inset showing the transitions in $\chi^{-1}$ vs *T*. $\chi_0$ is the background susceptibility that was subtracted from χ(*T*) to obtain a linear $\chi^{-1}$ vs *T* region. (b, c) *m* vs $\mu_0H$ and d*m*/d*H* vs $\mu_0H$ at various values of *T* for the Tb (b) and Dy (c) samples. The transition fields for the spin-flop-like process (inflection points for *m* vs $\mu_0H$ and peaks for d*m*/d*H* vs $\mu_0H$) are located by the black dashed lines.

Distinct from the nickelates discussed above, the Tb and Dy substituted samples exhibit clear signatures of MO before HPHT $O_2$ annealing (see Fig. 4). The χ(*T*) curve of the Yb sample (in Fig. S2) shows characteristics associated with the magnetic $LaYbO_3$ impurity [54]. The ZFC χ(*T*) curve of $La_{2.7}Tb_{0.3}Ni_2O_7$ ($La_{2.7}Dy_{0.3}Ni_2O_7$) has a shoulder at $T_N \approx 13.2$ K (10.7 K) and a peak at $T_N' \approx 4.8$ K (3.0 K). The *m*(*H*) curve of these two nickelates is reminiscent of a spin-flop transition for an antiferromagnet: at $T < T_N$, d*m*/d*H* experiences a rapid increase at the spin-flop point and then a decrease with increasing *H*; at $T > T_N$ (paramagnetic region), *m*(*H*) can be described by a Brillouin function with a monotonic decrease in d*m*/d*H* with *H*. Considering these are polycrystalline samples with random crystallite orientations, Figs. 4(b, c) basically reflect the spin-flop features of antiferromagnets. Therefore, we propose that it is likely that $La_{2.7}Tb_{0.3}Ni_2O_7$ and $La_{2.7}Dy_{0.3}Ni_2O_7$ undergo an AFM transition at $T_N$ and, possibly, another AFM transition at $T_N'$ based on the peak feature in χ(*T*). We don't exclude other possible sources of the anomalies at $T_N'$ such as another type of MO, a spin-glass transition [44], or spin rearrangement. Noting that $Tb_2O_3$, $Tb_4O_7$, $Dy_2O_3$, and NiO have AFM transitions at 2.42 K [55], 7.85 K [55], 1.2 K [56], and 523 K [57], respectively, it is unlikely that the MO transitions are from these impurities. Furthermore, if these impurity phases are present, they are only at levels below the detection limit of PXRD.

Assuming that the magnetic susceptibility is dominated by the contribution from the *Ln* ions, Curie-Weiss fitting yields the Curie-Weiss temperature of $T_{CW} \approx -11$ K and $\mu_{eff} \approx 9.73\ \mu_B$ for the Tb sample, and $T_{CW} \approx -5$ K and $\mu_{eff} \approx 10.58\ \mu_B$ for the Dy sample (see Table S1 in the Supporting Information). In both cases, the values obtained for $\mu_{eff}$ are in excellent agreement with the theoretical free ion values for $Tb^{3+}$ (9.72 $\mu_B$) or $Dy^{3+}$ (10.63 $\mu_B$) ions, indicating that the high-temperature paramagnetism primarily originates from localized $Tb^{3+}$ and $Dy^{3+}$ moments, which are apparently responsible for antiferromagnetic order below $T_N$ due to predominantly antiferromagnetic interactions between the $Ln^{3+}$ ions suggested by the negative sign of $T_{CW}$. Jiao *et al.* [43] synthesized $La_{3-x}Tb_xNi_2O_7$ (x = 0.1 – 0.5) powdered samples via solid state reaction and observed a broad maximum at ~4.5 K in $\chi(T)$ like the peak we found at $T_N$'. They suggested that this apparent MO could be due to magnetic exchange interactions between Tb and Ni ions [43]. In addition to this, Jiang *et al.* [44] prepared $La_{3-x}Tb_xNi_2O_7$ (x = 0.1 – 0.3) powders using the sol-gel method and observed a peak (shoulder) at ~4.5 K (~12 K) in $\chi(T)$ of their x = 0.3 sample, similar to what we have reported in this paper, but they attributed the peak to the formation of a spin-glass involving the Tb ions [44].

Although the detailed magnetic structures of the Tb- and Dy-substituted samples remain undetermined, we tentatively employ a simplified two-sublattice collinear antiferromagnetic model to obtain estimates for the relevant magnetic parameters. Using the Weiss molecular field model and assuming that the apparent AFM order below $T_N$ comes from the antiparallel alignment of two sublattices of $Ln^{3+}$ magnetic moments, we can estimate the exchange field ($H_E$), the anisotropy field ($H_A$), and the spin-flop to paramagnetic transition field ($H_{SF-P}$) at zero temperature from the Néel temperature $T_N$, $J$ of free $Tb^{3+}$ and $Dy^{3+}$ ions, and the spin-flop field ($\mu_0 H_{SF}$, approximately 1.1 T and 0.8 T for the Tb and Dy substituted samples, respectively) determined from the maximum of d$m$/d$H$ as seen in Fig. 4(b,c) [58, 59, 60, 61]. The following equations are used:

$$H_{SF} = (2H_E H_A - H_A^2)^{1/2} \qquad (1)$$

$$H_{SF-P} = 2H_E - H_A \qquad (2)$$

$$\mu_0 H_E = 3k_B T_N/[g_J \mu_B (J+1)] \qquad (3)$$

where $g_J = 3/2$ (4/3) for $Tb^{3+}$ ($Dy^{3+}$) and $k_B$ is the Boltzmann constant. By substituting measured values of $T_N$, $\mu_0 H_{SF}$, we found $\mu_0 H_E \approx 5.6$ T and 4.2 T, $\mu_0 H_A = 0.11$ T and 0.08 T, and $\mu_0 H_{SF-P}$ = 11.1 T and 8.4 T, for the Tb and Dy substituted samples, respectively. The small ratio of $H_A / H_E \approx 0.019$ (Tb) or 0.018 (Dy) for the two samples indicates that the exchange interaction dominates the magnetic ground state over the magnetic anisotropy and that the two AFM sublattices remain rigidly coupled, allowing the $Tb^{3+}$ ($Dy^{3+}$) magnetic moments to undergo a coherent rotation under an applied magnetic field through the spin flop transition.

### *3.4 Neutron scattering results for Tb and Dy substituted nickelates*

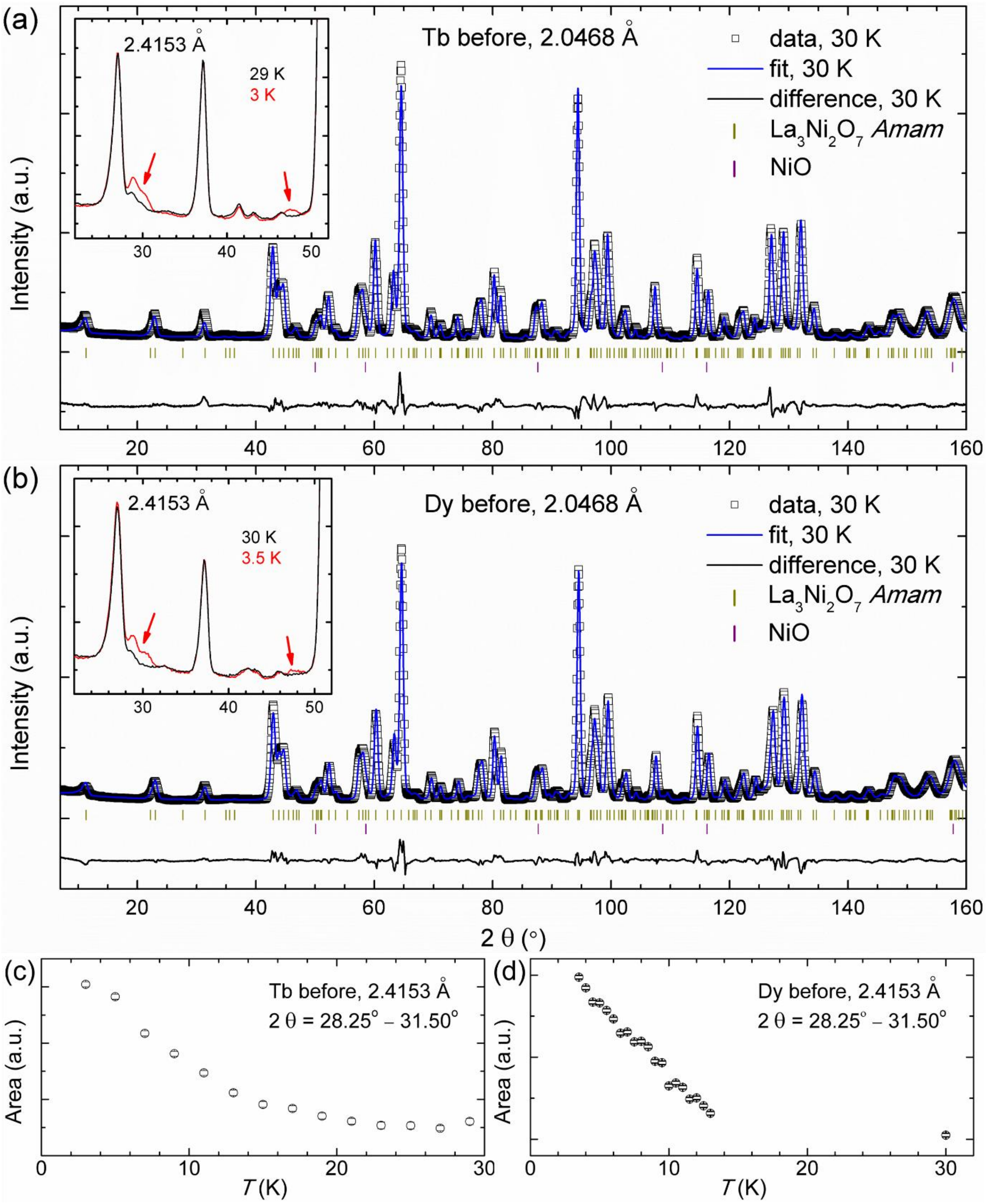

Figure 5. Neutron scattering for $La_{2.7}Tb_{0.3}Ni_2O_7$ and $La_{2.7}Dy_{0.3}Ni_2O_7$ at low temperatures before annealing. (a, b): Diffraction patterns measured at 2.0468 Å and Rietveld refinements confirming that the main phase adopts the *Amam* structure and the existence of ~1.1 wt% NiO impurity. No lanthanide-oxide structural or magnetic impurity phases could be detected. The insets show patterns at 2.4153 Å, at ~ 3 K and ~ 30 K, revealing the appearance of magnetic diffraction peaks (indicated by red arrows) that cannot be indexed to the *Amam* nuclear structures. (c, d): Temperature-dependent integrated areas of Gaussian-fitted magnetic double-peaks between 28.25° and 31.50°. The uncertainties in the integrated intensity are given as error bars. The full widths at half maximum, 1.21(1) and 1.42(1)°, for the two peaks of the doublet were fitted independently but as temperature invariant.

To further study the MO in $La_{2.7}Tb_{0.3}Ni_2O_7$ and $La_{2.7}Dy_{0.3}Ni_2O_7$, NPD experiments were performed in zero magnetic field and at low temperature with results shown in Fig. 5. At 30 K, neutron diffraction data obtained by the high-resolution diffractometer Echidna (2.0468 Å) for both samples can be well fitted with Rietveld refinements for the *Amam* $La_3Ni_2O_7$ phase and a minor NiO impurity phase. No evidence for $Tb_2O_3$, $Tb_4O_7$, or $Dy_2O_3$ impurities could be detected. In addition, very weak extra peaks near the (0 0 4) and (1 1 5) peaks of *Amam* $La_3Ni_2O_7$ were observed at lower temperatures. Using the high-intensity diffractometer Wombat and 2.4153 Å neutron beam, we observed enhancement of three extra peaks with cooling, as shown in Figs. 5 (a, b) insets, indicating the development of long-range MO. These peaks cannot be indexed by the nuclear structures of *Amam* $La_3Ni_2O_7$ and therefore suggest that the magnetic structure due to the respective Dy or Tb magnetic moments is incommensurate, although the paucity of peaks and weak intensities prohibit any further conclusions. It is therefore especially important to investigate if these magnetic peaks could be due to an impurity phase. The magnetic order of NiO occurs at much higher temperatures to be relevant. More importantly, there is no correspondence between these observed magnetic peak positions and the calculated structural or magnetic peak positions for $Dy_2O_3$, $Tb_2O_3$, or $Tb_4O_7$. The double-peak feature centered near 30° was fit with two Gaussian functions and the integrated double-peak area vs. temperature plots are given in Figs. 5 (c, d) for the Tb and Dy samples before annealing, showing a marked increase as the possible AFM order develops below 13 K and 11 K for the Tb and Dy samples, respectively. We note that these transition temperatures also do not correspond with the ordering temperatures of the above lanthanide oxides.

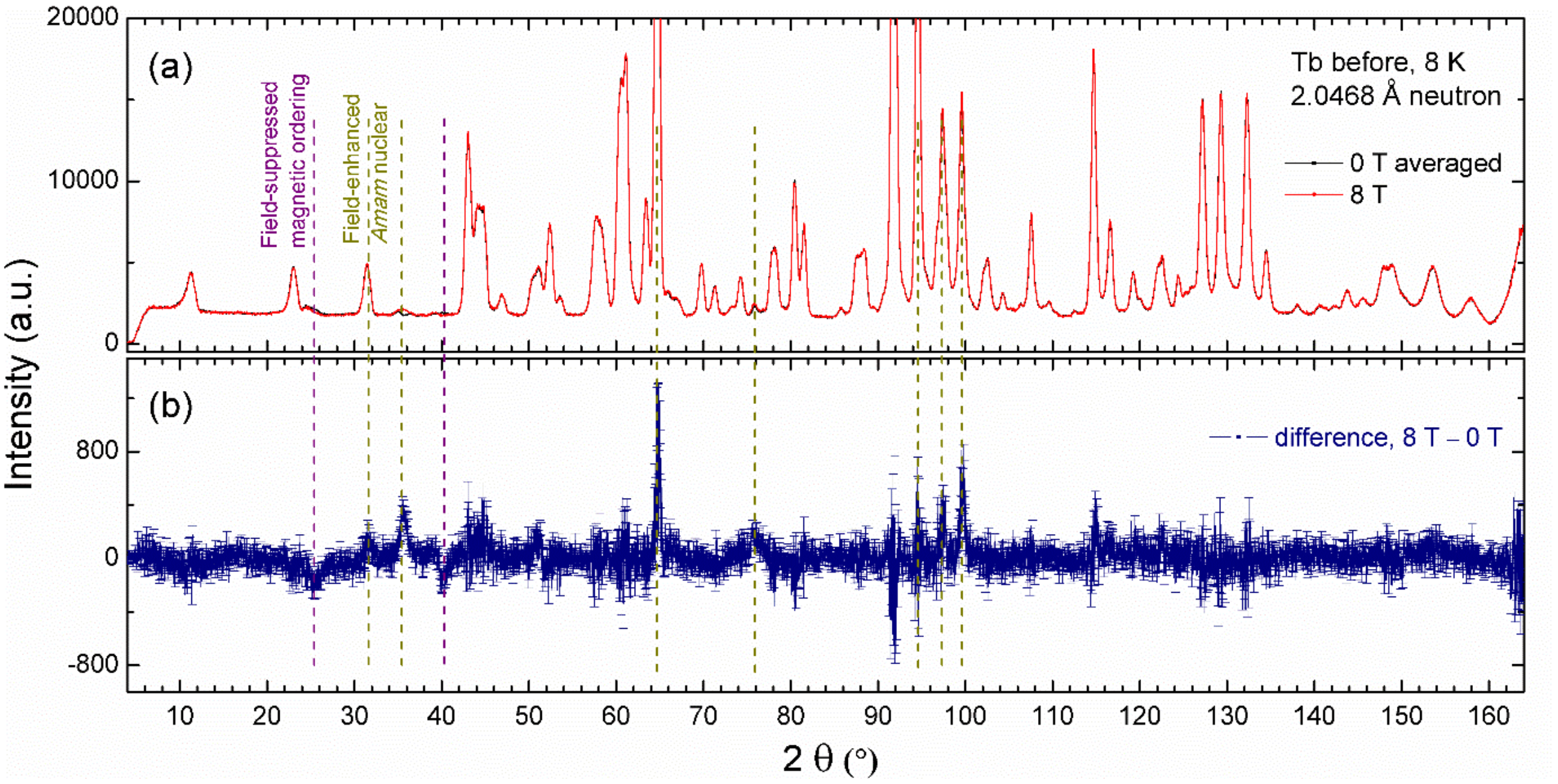

Figure 6. Magnetic field-dependent neutron powder diffraction patterns of $La_{2.7}Tb_{0.3}Ni_2O_7$ before annealing at 8 K. (a) Diffraction patterns collected at 0 and 8 T. The zero-field pattern is the average of measurements taken before application and after removal of the 8 T field. (b) Difference pattern obtained by subtracting the averaged zero-field pattern from the 8 T pattern. Negative and positive intensity differences indicate field-suppressed magnetic reflections and field-enhanced scattering at selected nuclear Bragg positions of the *Amam*-327 phase, respectively, as marked by the purple and brown dashed lines.

Applying a magnetic field of 8 T to $La_{2.7}Tb_{0.3}Ni_2O_7$ before annealing at 8 K, below $T_N$, suppresses the additional reflections attributed to the AFM ordering, which cannot be indexed by the nuclear *Amam*-327 structure, while enhancing the scattering intensity at selected nuclear Bragg positions of the 327 phase, as shown in Fig. 6. These changes are consistent with a field-induced reduction of the AFM component and the development of a uniform magnetization component as the magnetic moments increasingly align with the applied field. The latter produces magnetic scattering at the nuclear Bragg positions because it has the same translational periodicity as the crystal lattice. Both the suppression of the additional magnetic reflections and the enhancement of selected nuclear Bragg peaks are reversible upon removal of the magnetic field. No new Bragg reflections are observed upon application of the 8 T field. Taking together the detailed neutron and x-ray diffraction data and the magnetization results, our observations support the interpretation that the low-temperature magnetic ordering is antiferromagnetic and associated with the *Amam*-327 phase rather than an unidentified magnetic impurity phase. Definitive investigations of the magnetic structure and order will have to await the availability of single crystals.

The *a*, *b*, and *c* lattice parameters and the interlayer Ni-O1-Ni bond angle (O1: inner apical oxygen) $\theta_{Ni\text{-}O\text{-}Ni}$ of the Tb and Dy samples were obtained from the refined crystal structures using VESTA as summarized in Table 1, with corresponding values for $La_3Ni_2O_7$ synthesized in the same way [14]. Consistent with the PXRD results (see Fig. 2(b)), Tb and Dy substitutions lead to a contraction of the lattice along the *a*- and *c*-axis, with nearly no change along the *b*-axis. Notably, the *b* parameter barely changes from room temperature to 3 K, in contrast to the obvious decrease in the *a* and *c* parameters, revealed by our room-temperature and low-temperature neutron diffraction data. $La_{2.7}Tb_{0.3}Ni_2O_7$ and $La_{2.7}Dy_{0.3}Ni_2O_7$ have a smaller value of $\theta_{Ni\text{-}O\text{-}Ni}$ than the undoped sample, comparable to the angle of 164.2(5)° for $La_2SmNi_2O_7$ [31] and ~164.2° for $La_2PrNi_2O_7$ [24]. Given that the angle decreases with increasing Pr substitution concentrations [24] and is ~166.2° in $La_{2.7}Pr_{0.3}Ni_2O_7$ [24], Tb and Dy substitutions exhibit the strongest per-*Ln* suppression of $\theta_{Ni\text{-}O\text{-}Ni}$ investigated to date. Inspection of the refined crystal structures of the Tb and Dy samples reveals that the Ni-O1-Ni bond is bent along the *b*-axis, suggesting a scenario that upon Tb/Dy doping, the *Ln* contraction drives the reductions of the *a* and *c* parameters, whereas bending of the Ni–O1–Ni bond partially compensates the expected contraction along the *b*-axis, resulting in an almost unchanged *b* parameter.

Table 1. Crystal structure information for *Amam* $La_{2.7}Tb_{0.3}Ni_2O_7$ and $La_{2.7}Dy_{0.3}Ni_2O_7$, both determined from measurements on the Echidna spectrometer (neutron wavelength = 2.0468 Å), and $La_3Ni_2O_7$ [14] at low temperature and at ambient pressure.

| | $La_3Ni_2O_7$, 5 K | $La_{2.7}Tb_{0.3}Ni_2O_7$, 5 K | $La_{2.7}Dy_{0.3}Ni_2O_7$, 3.3 K |
|---|---|---|---|
| $\theta_{Ni\text{-}O\text{-}Ni}$ (°) | 168.6(11) | 163.9(4) | 163.4(4) |
| *a* (Å) | 5.39813(16) | 5.36270(17) | 5.36133(15) |
| *b* (Å) | 5.46493(18) | 5.45159(16) | 5.45319(14) |
| *c* (Å) | 20.48640(7) | 20.35240(70) | 20.32720(60) |
| Ni-O1 (Å) | 1.9641(19) | 1.985(3) | 2.0013(18) |
| Ni-O2 (Å) | 2.1560(15) | 2.214(5) | 2.191(4) |
| Ni-O3 (Å) | 2.006 (6) | 1.910(3) | 1.912(3) |
| Ni-O4 (Å) | 2.042(7) | 1.933(3) | 1.932(3) |

Unlike pressure, which reduces the orthorhombic distortion of $La_3Ni_2O_7$ and promotes superconductivity, Tb and Dy substitution enhances the in-plane lattice anisotropy, suggesting a structural trend away from the superconducting tetragonal phase. In addition, the reduced $\theta_{Ni-O-Ni}$ may further suppress superconductivity in these compounds. In an octahedral crystal field, the Ni-$e_g$ orbitals ($d_{z^2}$ and $d_{x^2-y^2}$) have higher energy than the $t_{2g}$ orbitals and are partially occupied, playing a central role in determining the electronic structure and the superconductivity in $La_3Ni_2O_7$ [12, 62]. The O1-2$p_z$ orbital mediates interlayer coupling by hybridizing with Ni-$d_{z^2}$ orbitals, thereby providing a hopping channel for interlayer electrons. A reduced $\theta_{Ni-O-Ni}$ in the Tb and Dy samples is expected to decrease this hopping probability [24], consistent with the higher electrical resistivity shown in Fig. 7(a), and to weaken the interlayer $d_{z^2}$ -derived electronic coupling that has been proposed to promote superconductivity [63, 64].

However, the observation that $La_2SmNi_2O_7$ [31] and $La_2PrNi_2O_7$ [24], despite having similarly reduced $\theta_{Ni-O-Ni}$, still exhibit superconductivity – with $La_2SmNi_2O_7$ even showing a higher $T_c$ than $La_3Ni_2O_7$ – highlights the importance of the in-plane Ni-$d_{x^2-y^2}$ orbitals [42, 65]. In fact, with Tb/Dy substitution, the in-plane Ni-O3 and Ni-O4 bond lengths shrink at the expense of elongated out-of-plane Ni-O1 and Ni-O2 (O2: outer apical oxygen) bond lengths as shown in Table 1, suggesting an enhanced in-plane (Ni-$d_{x^2-y^2}$ – O-$p_{x,\ y}$ – Ni-$d_{x^2-y^2}$) hybridization relative to the out-of-plane (Ni-$d_{z^2}$ – O-$p_z$ – Ni-$d_{z^2}$) hybridization. In this context, the high-pressure studies of $La_{2.7}Tb_{0.3}Ni_2O_7$ and $La_{2.7}Dy_{0.3}Ni_2O_7$ are motivated by multiple intertwined factors, including *Ln*-induced magnetic ordering, enhanced orthorhombicity, and modified orbital-selective hybridization between Ni-$d_{z^2}$ and Ni-$d_{x^2-y^2}$ states. Disentangling these intertwined degrees of freedom is essential for identifying the primary superconducting mechanism in $La_3Ni_2O_7$-based systems.

### *3.5 More metallic nickelates due to high-pressure high-temperature $O_2$ annealing*

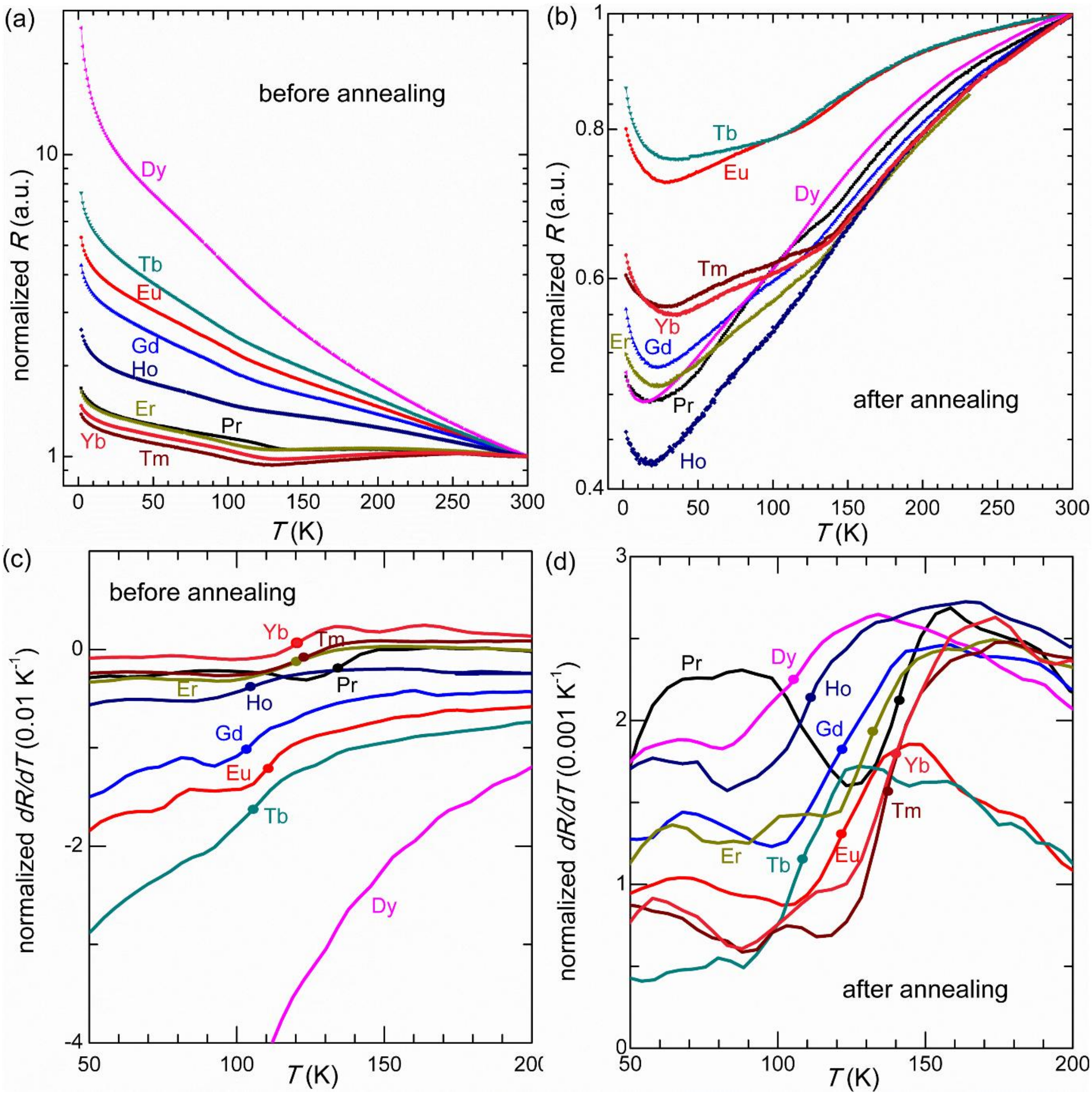

Figure 7. Normalized electrical resistivity of the nickelates with the signatures of density-wave transitions. *R* vs *T* before (panels a and c) and after (panels b and d) HPHT $O_2$ annealing. $T_{DW}$ is determined by the midpoint of the downturn in d*R*/d*T* upon cooling as indicated by solid circles in panels b and d).

Before HPHT $O_2$ annealing, the behavior of *R*(*T*) for all the samples was complex as shown in Fig. 7(a). At high temperature, the electrical resistivity (ρ) of the Pr, Er, Tm, and Yb substituted samples exhibits a weak *T* dependence, whereas ρ(*T*) of the Eu, Gd, Tb, Dy, and Ho substituted samples displays more semiconducting-like behavior; upon cooling, all of the samples exhibit a resistivity upturn below a characteristic temperature $T_{DW}$ which should correspond to the DW transition reported in $La_3Ni_2O_7$ [12, 14], $La_{3-x}Nd_xNi_2O_7$ [30], and $La_{3-x}Tb_xNi_2O_7$ [43]. To consistently determine the temperature $T_{DW}$, d*R*/d*T* plots were used (the method fails for the Dy sample) as shown in Fig. 7(c), which reveals that $T_{DW}$ varies between 104 K and 134 K across the series with the smallest values observed for the Gd, Tb, and Ho substituted samples. Also, the substitution of *Ln* ions, most notably Gd, Tb, and Ho, suppresses $T_{DW}$ of $La_3Ni_2O_7$ significantly from ~135–140 K to ~110 K, resembling the suppression of $T_{DW}$ with pressure [13, 66]. After annealing, all samples exhibit more metallic $R(T)$ behavior at high temperature and an increase in resistivity with decreasing $T$ below ~30 K. Notably, the DW transitions are still observable with slightly higher values of $T_{DW}$ than in the as-synthesized samples, suggesting that the DW state is robust against both *Ln* substitution and oxygen stoichiometry tuning, although its detailed nature may be modified.

HPHT $O_2$ annealing has been demonstrated to metallize $La_3Ni_2O_7$ and its *Ln* substituted relatives [32, 34, 43]. The electrical transport of RP nickelates is governed by the Ni-3*d* electrons whose degree of localization depends on the Ni valence, oxygen stoichiometry, and the geometry of the structure [67, 68, 69]. Oxygen annealing is generally believed to reduce O-vacancies mainly at the inner apical oxygen sites that bridge interlayer Ni-O octahedra [32]. The increase in oxygen content raises the nominal Ni valence, thereby increasing the charge carrier concentration [43]. Additionally, the reduction of O-vacancies suppresses the perturbation of the periodic Ni-site potential, electron traps, and Anderson localization of states near the Fermi level $E_F$ [32]. Although we have not been able to determine the oxygen stoichiometry experimentally, based on previous studies employing similar synthesis and annealing procedures [14, 32, 43], the oxygen content of our samples is estimated to be close to 6.9 before annealing and approach 7.0 after annealing. The enhanced electrical conductivity observed following HPHT oxygen treatment may therefore be associated with an increased occupancy of the O1 sites, although this interpretation remains to be experimentally verified. On the other hand, as discussed earlier, HPHT $O_2$ annealing resulted in a mixture of the *Amam* and secondary phases, potentially including semiconducting *Cmmm* 1313-$La_3Ni_2O_7$ [49] which may contribute to the observed resistivity increase below ~30 K in the annealed samples seen in Fig. 7(b).

### *3.6 Suppressed magnetic ordering in Tb and Dy substituted nickelates due to $O_2$ annealing*

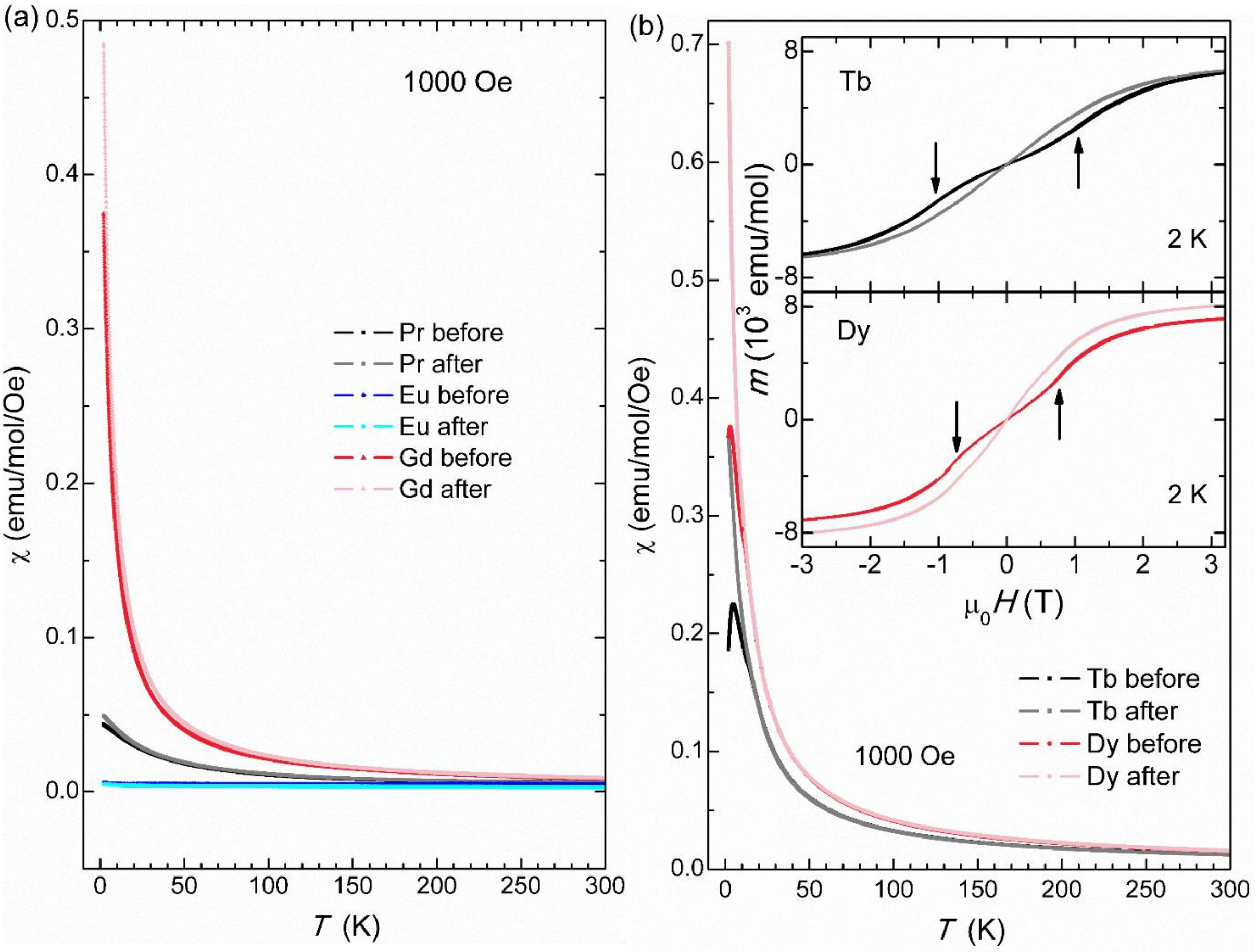

Figure 8. Magnetization of the nickelates before and after HPHT $O_2$ annealing. (a) $\chi(T)$ for the Pr, Eu, and Gd substituted samples. (b) $\chi(T)$ and $m(\mu_0 H)$ at 2 K (as insets) for the Tb and Dy substituted samples. The vertical arrows point to the inflection points observed in the as-prepared Tb and Dy substituted samples, which are strongly suppressed after annealing.

Although HPHT $O_2$ annealing introduces impurities and makes all samples more metallic, its impact on the magnetic properties is highly sample-dependent. As shown in Fig. 8(a) and Fig. S3, magnetism of the Pr-, Eu-, Gd-, Ho-, Er-, Tm-, and Yb-substituted compounds remain largely unchanged after annealing. In contrast, the magnetic signatures of the Tb and Dy substituted samples are strongly suppressed by HPHT $O_2$ annealing as seen in Fig. 8(b). The anomalies at $T_N$ and $T_N$' in $\chi(T)$ and the inflection points associated with the spin-flop-like transitions in $m(H)$ nearly disappear after annealing and are replaced by paramagnetic behavior of the $\chi(T)$ and $m(H)$ curves down to 2 K. These observations indicate a substantial suppression of the magnetic signatures associated with the low-temperature magnetic ordering, although they do not conclusively establish the complete disappearance of long-range magnetic order. As discussed previously, HPHT $O_2$ annealing can destabilize the *Amam* 2222-$La_3Ni_2O_7$ phase through the formation of secondary phases such as *Cmmm* 1313-$La_3Ni_2O_7$, *I*4/*mmm* 2222-$La_3Ni_2O_7$, and *Fmmm* 2222-$La_3Ni_2O_7$ [34]. We speculate that these structural defects may disrupt the magnetic exchange pathways responsible for the relatively fragile MO, characterized by low $T_N$, $T_N$' and $T_{CW}$, thereby weakening magnetic ordering in the Tb- and Dy-substituted compounds.

Interestingly, in contrast to our observations, Jiao *et al.* [43] reported that HPHT $O_2$ annealing enhanced the electrical conductivity of $La_{3-x}Tb_xNi_2O_7$ while largely preserving the low-temperature magnetic anomalies in $\chi(T)$. In their samples, the *Amam* structure was retained without substantial impurity-phase formation, whereas our annealing procedure resulted in appreciable structural modifications. This comparison suggests that structural changes may play an important role in the strong suppression of the magnetic anomalies observed in our samples, rather than oxygen incorporation alone. However, the relative contributions of changes in oxygen stoichiometry and crystal structure cannot be determined from the present measurements.

## 4. Discussion

In lanthanide intermetallic compounds, magnetic ordering temperatures and the suppression of $T_c$ in superconductors containing *Ln* impurities often scale with the dG factor of *Ln* ions because the strength of the RKKY interaction that leads to magnetic order is proportional to the dG factor [70, 71, 72, 73]. Among the $La_{3-x}Ln_xNi_2O_7$ compounds studied here, $Gd^{3+}$ has the largest dG factor (15.75), followed by $Tb^{3+}$ (10.50) and $Dy^{3+}$ (7.08); however, $La_{2.7}Gd_{0.3}Ni_2O_7$ remains paramagnetic down to 2 K whereas $La_{2.7}Tb_{0.3}Ni_2O_7$ and $La_{2.7}Dy_{0.3}Ni_2O_7$ develop magnetic order below ~13 K and ~11 K, respectively. The departure from the simple dG scaling suggests that additional factors beyond the dG factor itself are important in determining the magnetic ground state, although it does not rule out RKKY interactions. One possible origin of this deviation is the distinct influence of crystal-field effects on the different *Ln* ions. While $Tb^{3+}$ and $Dy^{3+}$ possess nonzero orbital angular momentum $L$ and may exhibit substantial crystal-field-induced magnetic anisotropy, $Gd^{3+}$ has an $L = 0$ ground state insensitive to crystal fields. Such differences in magnetic anisotropy may influence the stability of magnetic order and contribute to the contrasting magnetic behavior of these compounds. The ratios of the ordering temperatures ($T_N$ ratio 13.2/10.7 ≈ 1.23, $T_N$' ratio 4.8/3.0 ≈ 1.6) in the Tb and Dy compounds are also broadly comparable to the ratio of their dG factors (1.48), indicating that the dG scaling may still play a role in determining the magnetic ordering temperatures, even though it does not establish the underlying exchange mechanism.

The transport properties of these materials provide additional insights into the magnetic ordering. Prior to annealing, all compounds exhibit bad-metal-like transport traits. The MO of the Tb- and Dy-substituted samples produce little or no signatures in $R(T)$, while the DW transition strongly affects transport but is essentially invisible in $\chi(T)$. This contrast suggests that the MO of the Tb/Dy moments has a relatively weak influence on the electrical transport properties. Given the highly localized nature of the 4*f* orbitals in heavy lanthanides, indirect superexchange pathways involving oxygen ions may also contribute to the observed MO, alongside possible RKKY interactions. $La_2PrNi_2O_7$ [24], $La_{2.1}Nd_{0.9}Ni_2O_7$ [30], and $La_2SmNi_2O_7$ [31] exhibit pressure-induced superconductivity with $P_c$ and $T_c$ comparable to those of $La_3Ni_2O_7$. The apparent insensitivity of $T_c$ to substantial *Ln* substitution is reminiscent of $LnBa_2Cu_3O_{7-\delta}$, where superconductivity generally persists despite low-temperature AFM ordering of the *Ln* moments [74, 75, 76, 77]. Exceptionally, the nonsuperconducting $PrBa_2Cu_3O_{7-\delta}$ exhibits AFM at an unusually high temperature of ~17 K, which has been discussed in connection with Pr-4*f*–O-2*p* hybridization and superexchange interactions [75, 78, 79, 80, 81, 82, 83, 84]. By analogy, superexchange pathways involving Tb/Dy and oxygen may contribute to the MO in $La_{2.7}Tb_{0.3}Ni_2O_7$ and $La_{2.7}Dy_{0.3}Ni_2O_7$.

The observation of density-wave-like transport anomalies and additional low-temperature magnetic order in $La_{2.7}Tb_{0.3}Ni_2O_7$ and $La_{2.7}Dy_{0.3}Ni_2O_7$ distinguishes them from previously studied *Ln*-substituted $La_3Ni_2O_7$ systems, in which comparable additional low-temperature magnetic ordering has not been established. Determining whether superconductivity emerges under pressure in these two materials, and how $P_c$ and $T_c$ evolve in the presence of magnetic order, represents an important direction for future work. Conversely, if superconductivity is absent, investigating whether and how a relatively low-temperature magnetic order influences or suppresses the superconducting state may provide valuable insight into the pairing mechanism of $La_3Ni_2O_7$-based superconductivity.

## 5. Concluding remarks

Bilayer Ruddlesden-Popper nickelates $La_2PrNi_2O_7$ and $La_{2.7}Ln_{0.3}Ni_2O_7$ (*Ln* = Eu, Gd, Tb, Dy) polycrystalline samples in the *Amam* structure were successfully synthesized via the sol-gel method. In contrast, 10% substitution with Ho, Er, Tm, and Yb did not yield single-phase *Amam* samples. While $La_2PrNi_2O_7$ and $La_{2.7}Ln_{0.3}Ni_2O_7$ (*Ln* = Gd, and probably Ho, Er, Tm, Yb) are paramagnetic and $La_{2.7}Eu_{0.3}Ni_2O_7$ is essentially nonmagnetic, distinct anomalies in magnetic susceptibility, spin-flop features in magnetization, and additional magnetic reflections in neutron powder diffraction provide evidence for long-range magnetic order in $La_{2.7}Tb_{0.3}Ni_2O_7$ and $La_{2.7}Dy_{0.3}Ni_2O_7$ below ~ 13 K and 11 K, respectively. The magnetic structures and ordering mechanism remain unresolved, although antiferromagnetic ordering is a plausible scenario.

HPHT $O_2$ annealing introduces impurity phases and produces more metallic-like transport behavior while preserving density-wave transitions as detected in the electrical resistivity measurements but strongly suppresses the magnetic anomalies associated with long-range magnetic order in the Tb- and Dy- substituted nickelates. The accompanying structural modifications may play an important role in this suppression, although contributions from changes in oxygen stoichiometry cannot be excluded. The presence of density-wave order and apparent long-range magnetic order in Tb- and Dy-substituted $La_3Ni_2O_7$ makes these compounds promising platforms for future investigations of the interplay among density-wave instabilities, magnetic order, and pressure-induced superconductivity. Whether the magnetic order persists under pressure and how it might interact with superconductivity remain open questions.

## Acknowledgements

Research at the University of California San Diego was supported by the U.S. Department of Energy (DOE) Basic Energy Sciences under Grant No. DE-FG02-04ER46105 (crystal growth) and the National Nuclear Security Administration (NNSA) under the Stewardship Science Academic Alliance Program through the U.S. DOE under Grant No. DE-NA0004086 (measurements at low temperature and in magnetic field). Research at the Federal University of ABC and the University of São Paulo was supported by Fundação de Amparo à Pesquisa do Estado de São Paulo (FAPESP) under Grant Numbers 2024/00989-7, 2024/18240-2, 2024/01816-9, 2022/10874-7, and 2013/07296-2, and Conselho Nacional de Desenvolvimento Científico e Tecnológico (CNPq) under Grant Number 304614/2023-8. (high-pressure high-temperature oxygen annealing and x-ray characterization). Research at Australian Nuclear Science and Technology Organisation was supported by the Australian Research Council under Discovery Project DP230100558 (neutron scattering). FA and RFJ acknowledge Paulo S. Martins for technical assistance. We benefitted from informative discussions with our colleague Prof. Alex Fraño. The identification of any commercial product or trade name does not imply endorsement or recommendation by the National Institute of Standards and Technology.

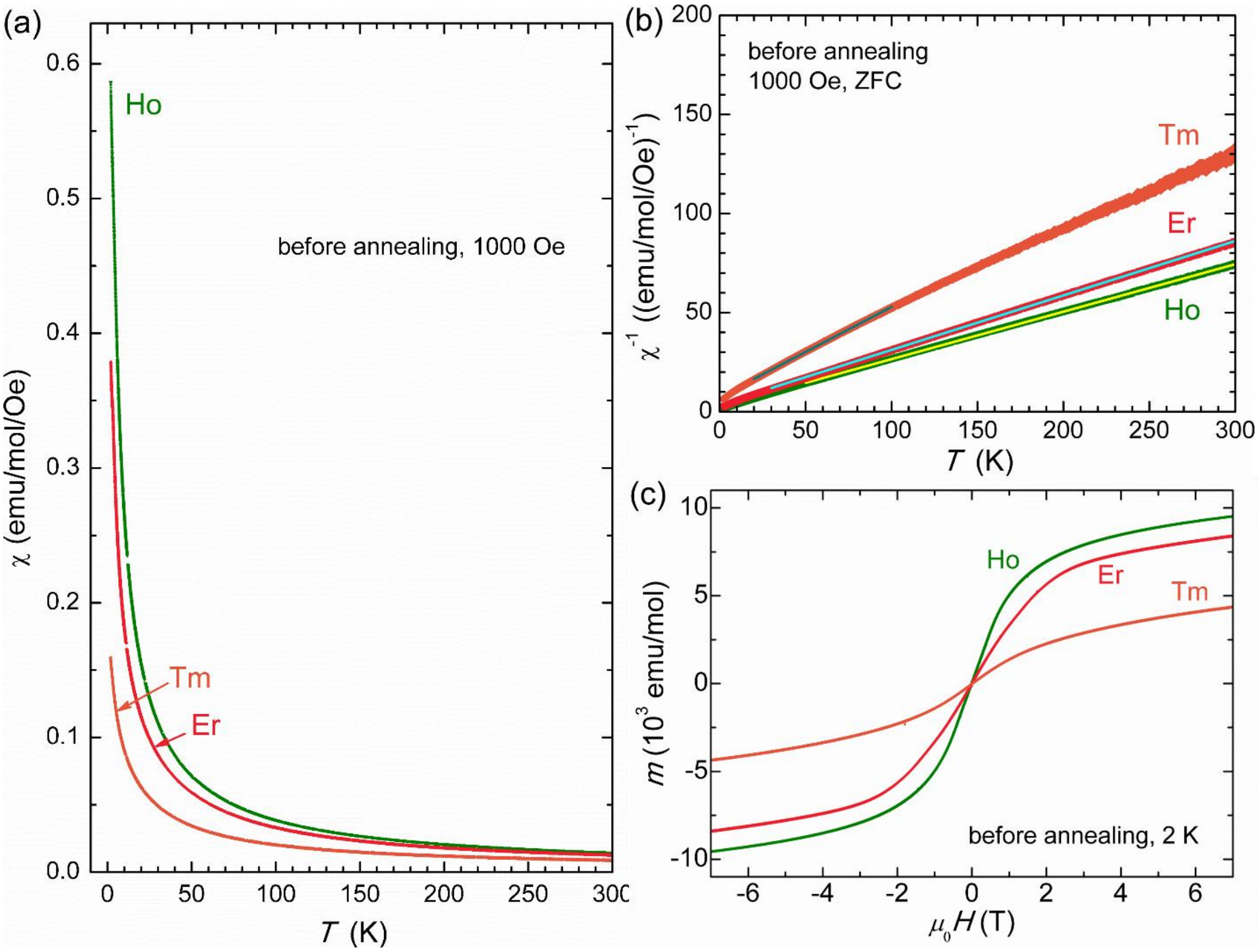

Figure S1. Magnetization of the Ho, Er, and Tm substituted samples in magnetic fields and at low temperatures before annealing. (a) $\chi$ vs $T$. (b) Background subtracted $\chi^{-1}$ vs $T$ with linear fitting in different $T$ ranges. (c) $m$ vs $\mu_0 H$ at 2 K. Note: 1 emu = $10^{-3}$ A m$^2$, and 1 Oe = (1000/4π) A/m.

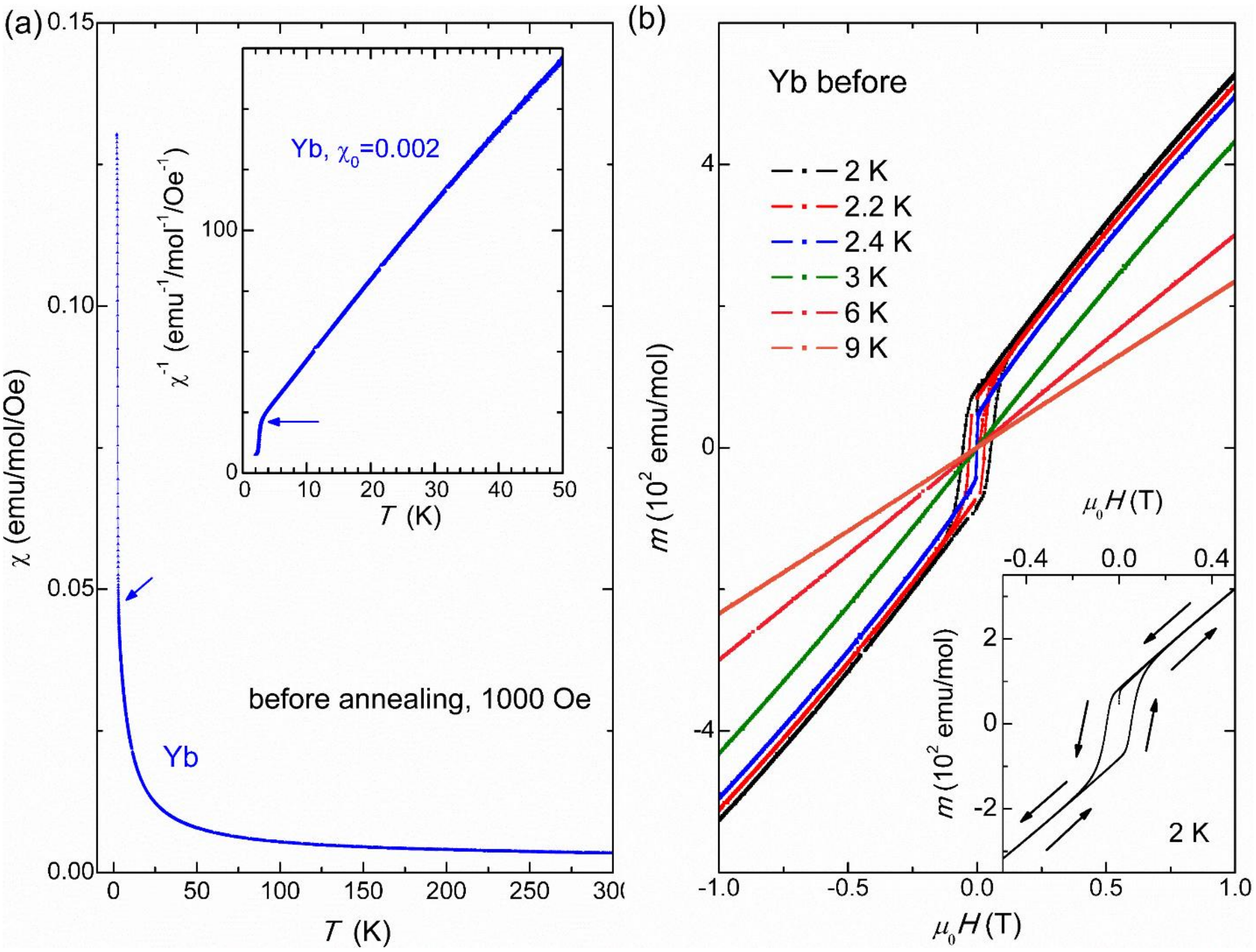

Figure S2. Magnetization of the Yb substituted samples before annealing in magnetic fields and at low temperatures. (a) $\chi$ vs $T$ for with an arrow pointing to the kink where magnetic ordering occurs, with an inset showing magnetic ordering transitions at low temperature as shown in $\chi^{-1}$ vs $T$. $\chi_0$ is the background susceptibility that was subtracted from $\chi(T)$ to obtain a linear $\chi^{-1}$ vs $T$ region. (b) $m$ vs $\mu_0H$ at various values of $T$ with an inset showing the enlarged ferromagnetic loop at 2 K. The kink and the loop should be associated with the impurity phase $LaYbO_3$ in the sample.

Table S1. Curie–Weiss fitting ranges and parameters (Curie-Weiss temperature $T_{CW}$, effective magnetic moment $\mu_{eff}$) after background ($\chi_0$) subtraction, together with theoretical free-ion effective magnetic moments, for the *Ln* substituted nickelate samples.

| Substitution | $T_{CW}$ (K) | $\mu_{eff}$ ($\mu_B$) | $\mu_{eff, theoretical, free}$ ($\mu_B$) | $\chi_0$ | Fitting range (K) |
|---|---|---|---|---|---|
| Pr | -18.51 | 3.06 | 3.58 | 0.001 | 30 – 140 |
| Gd | -5.65 | 7.55 | 7.94 | 0.001 | 2 – 300 |
| Tb | -11.1 | 9.73 | 9.72 | 0.001 | 30 – 300 |
| Dy | -4.61 | 10.58 | 10.63 | 0.001 | 30 – 150 |
| Ho | -10.64 | 10.55 | 10.60 | 0.001 | 50 – 300 |
| Tm | -16.3 | 7.66 | 7.57 | 0.001 | 20 – 100 |
| Er | -13.32 | 9.85 | 9.59 | 0.001 | 30 – 300 |
| Yb | -11.58 | 3.13 | 4.54 | 0.002 | 20 – 100 |

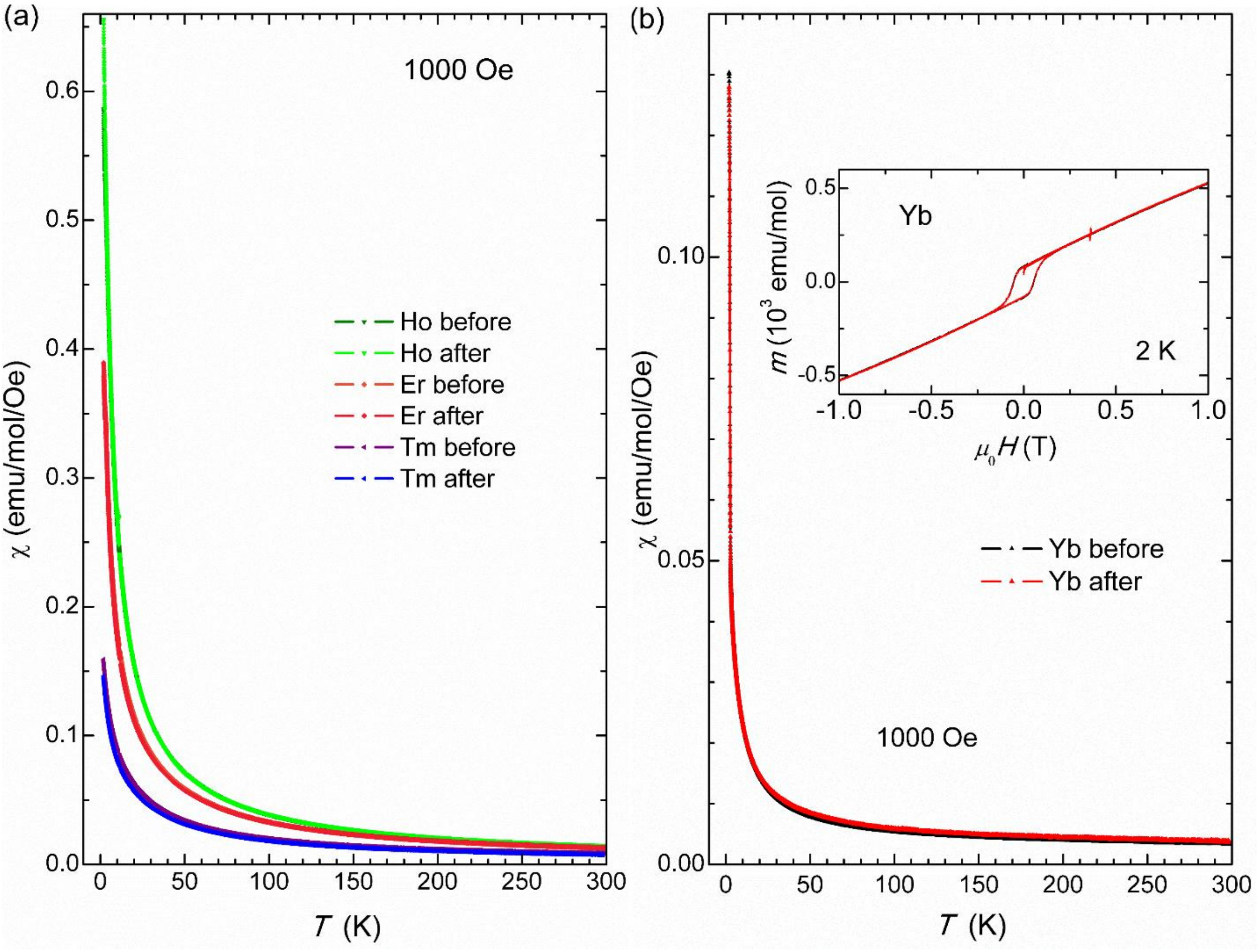

Figure S3. Magnetization of the Ho, Er, Tm, and Yb substituted samples before and after HPHT $O_2$ annealing. (a) $\chi(T)$ for the Ho, Er, and Tm substituted samples. (b) $\chi(T)$ and $m(\mu_0 H)$ at 2 K (as an inset) for the Yb substituted sample.

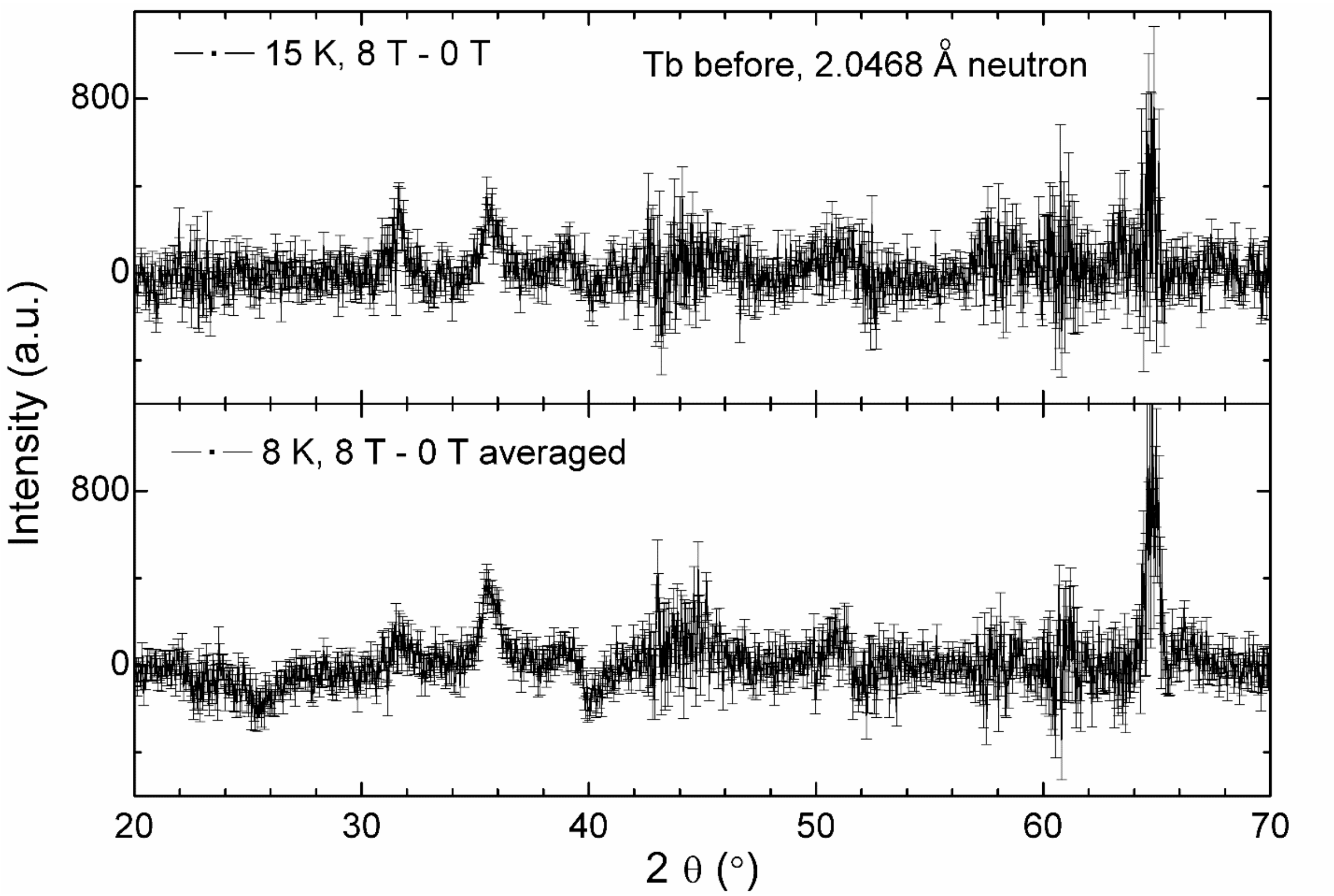

Figure S4. Difference neutron powder diffraction patterns of $La_{2.7}Tb_{0.3}Ni_2O_7$ before annealing at 8 K (lower panel) and 15 K (upper panel). The difference patterns were obtained by subtracting the zero-field diffraction patterns from those collected under an applied magnetic field of 8 T. For 8 K, the zero-field reference pattern was obtained by averaging the patterns collected before application and after removal of the 8 T field to improve the signal-to-noise ratio.